\documentclass[12pt,prd]{revtex4}

\usepackage{amsmath}
\usepackage{amssymb}
\usepackage[dvips]{graphicx,color}

\begin{document}

\newcommand{\VEV}[1]{\left\langle #1\right\rangle}
\newcommand{\wt}{\widetilde}
\newcommand{\wh}{\widehat}
\newcommand{\p}{\partial}
\newcommand{\nn}{\nonumber}
\newcommand{\tr}{\mathop{\rm tr}\nolimits}
\newcommand{\Tr}{\mathop{\rm Tr}\nolimits}
\newcommand{\diag}{\mathop{\rm diag}}
\newcommand{\cL}{{\mathcal L}}
\newcommand{\cO}{{\mathcal O}}
\newcommand{\cA}{{\mathcal A}}
\newcommand{\cF}{{\mathcal F}}
\newcommand{\cI}{{\mathcal I}}
\newcommand{\cD}{{\mathcal D}}
\newcommand{\cjpsi}{c}
\newcommand{\crho}{c_{\rho}}
\newcommand{\abs}[1]{\left| #1\right|}
\newcommand{\drv}[2]{\frac{d #1}{d #2}}
\newcommand{\Drv}[2]{\frac{\p #1}{\p #2}}
\newcommand{\MeV}{\;\text{MeV}}
\newcommand{\GeV}{\;\text{GeV}}
\newcommand{\fm}{\;\text{fm}}
\newcommand{\bra}[1]{\left\langle #1\right\vert}
\newcommand{\ket}[1]{\left\vert #1\right\rangle}
\newcommand{\braket}[2]{\langle #1\vert #2\rangle}
\newcommand{\cl}{{\rm cl}}
\newcommand{\ds}{\displaystyle}
\newcommand{\SCSnew}{S_{\rm CS}^{\rm new}}
\newcommand{\veps}{\varepsilon}
\newcommand{\bVEV}[1]{\bigl\langle #1\bigr\rangle}
\newcommand{\Exp}{\mbox{Exp}}
\newcommand{\rhoVEV}[1]{\VEV{#1}_{\!\rho}}

\hfill\parbox{4cm}
{
 KUNS-2240\\
 YITP-09-72\\
}

\title{
 Melting Spectral Functions of the Scalar and Vector Mesons\\
 in a Holographic QCD Model
}

\author{Mitsutoshi Fujita}
\email{mfujita@gauge.scphys.kyoto-u.ac.jp}
\affiliation{Department of Physics, Kyoto University, Kyoto 606-8502, Japan}

\author{Kenji Fukushima}
\email{fuku@yukawa.kyoto-u.ac.jp}
\affiliation{Yukawa Institute for Theoretical Physics, Kyoto University,
         Kyoto 606-8502, Japan}

\author{Toru Kikuchi}
\email{kikuchi@gauge.scphys.kyoto-u.ac.jp}
\affiliation{Department of Physics, Kyoto University, Kyoto 606-8502, Japan}

\author{Tatsuhiro Misumi}
\email{misumi@yukawa.kyoto-u.ac.jp}
\affiliation{Yukawa Institute for Theoretical Physics, Kyoto University,
         Kyoto 606-8502, Japan}

\author{Masaki Murata}
\email{masaki@yukawa.kyoto-u.ac.jp}
\affiliation{Yukawa Institute for Theoretical Physics, Kyoto University,
         Kyoto 606-8502, Japan}

\begin{abstract}
 We investigate the finite-temperature spectral functions of heavy
 quarkonia by using the soft-wall AdS/QCD model.
 We discuss the scalar, the pseudo-scalar, the vector, and the
 axial-vector mesons and compare their qualitative features of the
 melting temperature and growing width.
 We find that the axial-vector meson melts earlier than the vector
 meson, while there appears only a slight difference between the
 scalar and pseudo-scalar mesons which also melt earlier than the
 vector meson.
\end{abstract}

\maketitle


\section{Introduction}
\label{sec:Intro}

The strongly-correlated quark-gluon plasma (sQGP), which is hot and
dense matter out of quarks and gluons created at the Relativistic
Heavy-Ion Collider (RHIC) in the Brookhaven National Laboratory (BNL),
have been attracting a great deal of interest in its intrinsic
non-perturbative
properties~\cite{Gyulassy:2004zy,Shuryak:2004cy,RHIC}.  Although there
is no systematic way to study such a non-perturbative system at strong
coupling, a powerful technique has developed recently based on the
gauge/string
correspondence~\cite{Maldacena:1997re,Gubser:1998bc,Witten:1998qj}.
The idea is that one can treat the strong-coupling regime in the gauge
field theory on the boundary by solving the weak-coupling string
theory (or classical one in the large $N_c$ and large 't~Hooft
coupling limit) in the bulk Anti de~Sitter (AdS) space.

A well-known example of successful application of the gauge/string
duality to sQGP physics is the exact computation of the shear
viscosity to the entropy density ratio, i.e.\
$\eta/s=\hbar/(4\pi k_B)$ in an $\mathcal{N}=4$ supersymmetric
Yang-Mills
plasma~\cite{Policastro:2001yc,Buchel:2003tz,Kovtun:2004de,Kapusta:2008ng}.
This value of $\eta/s$ is much smaller than any observation in reality
except for the heavy-ion collisions;  the hydrodynamic model studies
imply that $\eta/s$ of QCD matter is as small as suggested by the
string theory~\cite{Teaney:2003kp,Romatschke:2007mq}.  Besides, it is
conjectured that $\eta/s=\hbar/(4\pi k_B)$ might be a universal lower
bound and applied to strong-coupling QCD as well as supersymmetric
models.

The smallness of $\eta/s$ is an important indication of the sQGP
because a larger reaction cross-section leads to a smaller $\eta$ in
gaseous states.  Actually perturbative QCD calculations cannot give
an account for small $\eta/s$ in the weak-coupling
regime~\cite{Hosoya:1983xm,Arnold:2000dr,Aarts:2002cc}.  The
Monte-Carlo simulation of QCD on the lattice is a powerful instrument
to look into the non-perturbative strong-coupling regime.  It is still
difficult to estimate $\eta$ in fully dynamical simulations with
quarks, but the (quenched) results so far are not inconsistent with
the string theory estimate~\cite{Nakamura:2004sy,Meyer:2007ic}.

Another important indication to the sQGP is the in-medium property of
heavy quarkonia such as $J/\psi$.  In the recent lattice QCD
simulations the $J/\psi$ spectral functions (SPFs) both above and
below $T_c$ have been successfully constructed by means of the maximum
entropy method (MEM)~\cite{Asakawa:2000tr}, which has revealed that
the mesonic correlation (a peak in the SPF) survives even above twice
of
$T_c$~\cite{Umeda:2000ym,Asakawa:2003re,Datta:2003ww,Aarts:2006nr,%
Umeda:2007hy}.  It is, however, a non-trivial question how to explain
such a high melting temperature for $J/\psi$ in a conventional way
using the non-relativistic model with the Debye screened
potential~\cite{Matsui:1986dk,Karsch:1987pv,Lucha:1991vn,Wong:2004zr,%
Mocsy:2007yj}.  We have not yet reached a full consensus on the
interpretation of the $J/\psi$ SPFs above $T_c$, though there are many
theoretical efforts.  Our present aim is to investigate this question
using the gauge/string duality along the same line as our previous
work~\cite{Fujita:2009wc}.

In Ref.~\cite{Fujita:2009wc} we calculated the SPFs in the vector
channel assuming that the heavy-quark sector is decoupled from
others.  In this work we will extend our analysis to other channels;
scalar, pseudo-scalar, and axial-vector mesons, namely, $\chi_{c0}$,
$\eta_c$, and $\chi_{c1}$.  Since the interpretation of $\chi_{c0}$
(i.e.\ whether it melts or not above $T_c$) was
controversial~\cite{Datta:2003ww,Aarts:2006nr,Umeda:2007hy}, it is
important to clarify whether $c\bar{c}$ states in all these channels
melt at $T>T_c$ and, if so, when they melt.  Under the situation that
the MEM construction of the SPFs in lattice QCD simulations are still
difficult for all these channels, it is valuable to take advantage of
the holographic QCD model to see what spectral shape would transpire
in the strongly-coupling system.  In this work we will use the
soft-wall AdS/QCD model~\cite{Da Rold:2005zs,Karch:2006pv} (see
Refs.~\cite{Erlich:2005qh,Ghoroku:2005vt} for related works).
Although the SPFs at finite temperature and density have been
discussed by means of the D3/D7 setup~\cite{Karch:2002sh}, it is not
straightforward to deal with the heavy-meson SPFs in the D3/D7 model.
This is because the only energy scale in this model is fixed by the
pion decay constant and the SPFs are given as a function of not $T$
and $M_q$ independently but only $T/M_q$ where $M_q$ is the quark
mass~\cite{Mateos:2006nu,Erdmenger:2007cm}.  The soft-wall model, in
contrast, has one more phenomenological parameter, $c$, which is fixed
by the meson spectrum.

One of the important features in the soft-wall AdS/QCD model is that a
parameter in the bi-fundamental scalar sector controls the chiral
symmetry breaking, that is, the chiral condensate.  Interestingly
enough, the chiral condensate is uniquely determined from the infrared
(IR) boundary condition that is specified by a smooth function in the
soft-wall model.  Therefore, in this model, there is no ambiguity in
the IR limit in the case at finite $T$ (see
Refs.~\cite{Ghoroku:2005kg,Kim:2007rt} for holographic approaches to
finite-$T$ mesons).  The regular IR boundary condition allows us to
evaluate the Minkowskian
correlator~\cite{Son:2002sd,Policastro:2002se,Teaney:2006nc} from
which we can compute the SPFs.  We will find, in view of the resultant
SPFs, that the axial-vector states melt faster than the vector ones.
We will then clarify its origin in the chiral symmetry breaking due to
the scalar field which yields a difference between the vector and
axial-vector equations of motion.  On the other hand, the scalar and
pseudo-scalar dissociation temperatures are almost the same;
$T\simeq T_c$.  Finally, before closing this paper, we shall take a
closer look at the vector channel.


\section{Soft-wall model}
\label{sec:model}

The principle to construct the AdS/QCD model is the bulk/boundary
correspondence or UV/IR relation.   The generating functional in the
gauge field theory is equivalent to the exponential of an on-shell
action in the gravity theory (GKP-W
relation~\cite{Gubser:1998bc,Witten:1998qj}),
\begin{equation}
 Z[\phi_0] = \VEV{e^{i\int dx\,\phi_0(x)\,\cO(x)}}_{\text{gauge}}
  =  e^{iS_{\text{gravity}}[\phi_0]} ,
\label{eq:GKPW}
\end{equation}
where $Z[\phi_0]$ is the generating functional with the source
$\phi_0$ coupled with an operator $\cO(x)$ and $S_{\text{gravity}}$ is
an on-shell action with the boundary condition $\phi\to\phi_0$ at
the UV boundary where the gauge theory resides.

The AdS/QCD models are five-dimensional field theories designed to
describe QCD properties through the bulk/boundary
correspondence~\cite{Erlich:2005qh}.  The essential ingredients of the
AdS/QCD model are the AdS space with an IR cutoff (i.e.\ wall) that
translates into a typical energy scale in QCD,
the $\mathrm{U_L}(2)\times\mathrm{U_R}(2)$ vector fields, $A_L$,
$A_R$, and the bi-fundamental scalar field $X$.  The vacuum
expectation value of $X$ is responsible for the explicit and
spontaneous chiral symmetry breaking.  The soft-wall model is defined
by the following action~\cite{Karch:2006pv};
\begin{align}
 S &= \int d^5x\, e^{-cz^2}\sqrt{-g}\,\cL
\label{eq:5daction}\\
 \cL &= \tr\biggr[ -|DX|^2 + \frac3{L^2}|X|^2
 -\frac{1}{4g_5^2} \bigl( g^{MN}g^{PQ}F_{L,MP}F_{L,NQ} +
 g^{MN}g^{PQ}F_{R,MP}F_{R,NQ} \bigr) \biggl] ,
\label{eq:L}
\end{align}
where $X=X^at^a$ with $t^a$ being the generator of $U(2)$ and
$D_M X=\partial_M X+iA_{L,M}X-iXA_{R,M}$.  Here $M=x_0,x_1,x_2,x_3,z$
and $g_5$ are the indices for the five-dimensional coordinates and
gauge coupling, respectively.  We note that $g_5^2 = 24\pi^2L/N_c$ is
concluded by matching~\cite{Karch:2006pv}.  We also use the Greek
index $\mu=x_0,x_1,x_2,x_3$ to refer to the four-dimensional
coordinates.  The model parameter, $c$, characterizes the wall
location;  since the contribution from the IR region
$z\gtrsim 1/\sqrt{c}$ is suppressed by $e^{-cz^2}$, it represents a
potential with the wall providing a typical QCD scale.  The background
geometry is specified as the AdS metric as
\begin{equation}
 g_{MN}dx^M dx^N = \frac{L^2}{z^2}\bigl( -dt^2+d\vec{x}^2+dz^2 \bigr).
\label{eq:AdSmetric}
\end{equation} 

It should be mentioned that the vector meson mass spectra at $T=0$ are
quantized by the normalizability condition and given as the following
Regge trajectory~\cite{Karch:2006pv},
\begin{equation}
 m_n^2 = 4\,c\,n ,
\label{eq:Reggerho}
\end{equation}
where $n$ is the radial excitation number.  Then we can determine $c$
by fitting the above relation to the vector meson spectra;
$\rho(770)$, $\rho(1450)$, $\rho(1700)$, etc.  If we take
$m_{\rho}=0.77\GeV$ for $n=1$, we have $c=0.77^2/4=0.148\GeV^2$, while
we will later find that the spectral peak is slightly shifted from
Eq.~\eqref{eq:Reggerho} and will fix $c=0.151\GeV^2$ to fit the peak
position with the mass.

Now that we fix the model parameter $c$, let us consider the model at
finite temperature.  Here we shall introduce the following background,
which is called the AdS blackhole (AdSBH),
\begin{equation}
 g_{MN}dx^M dx^N = \frac{L^2}{z^2}\biggl( -f(z)dt^2 + d\vec{x}^2
  + \frac{1}{f(z)}dz^2 \biggr) , 
\label{eq:BHmetric}
\end{equation} 
with $f(z)= 1-z^4/z_h^4$ where the horizon is related to the Hawking
temperature that is interpreted as the system temperature of dual QCD
as $z_h=1/(\pi T)$.  It is known that the AdSBH is unstable at low
temperature, and thus the Hawking-Page-type transition occurs at a
critical temperature, $T_c=0.492\sqrt{c}$.  This is a first-order
phase transition from the AdSBH to the thermal AdS
metric~\cite{Andreev:2006eh,Herzog:2006ra} and can be identified as
the color confinement-deconfinement phase transition.

In the soft-wall model one can introduce the chiral symmetry breaking
explicitly (i.e.\ quark mass) and spontaneously (i.e.\ chiral
condensate) through the bi-fundamental scalar field $X$, which is
decomposed as $X(x,z)=e^{2i\Pi(x,z)}[X_0(z)+S(x,z)]$ where $x$ refers
to four-dimensional coordinates only and $X_0$ is a constant
background with respect to $x$.  The fluctuations, $S$ and $\Pi$,
represent the scalar and pseudo-scalar fields.  In the case when quark
masses are degenerated, $X_0(z)$ is proportional to unity in flavor
space and satisfies the following equation of motion,
\begin{equation}
 X_0''(z) + \biggl( -2cz + \frac{f-4}{zf} \biggr) X_0'(z)
  + \frac{3}{z^2f}X_0(z) = 0 ,
\label{eq:eomX0}
\end{equation}
where the prime stands for the derivative with respect to $z$.  From
this differential equation we find that in the vicinity of $z=0$ the
general solution behaves as
\begin{equation}
 L^{3/2}X_0(z) \sim \frac{1}{2}( M_q z + \Sigma z^3 ) ,
\label{eq:X0-ini}
\end{equation}
where, according to the dictionary of bulk/boundary correspondence,
the parameters $M_q$ and $\Sigma$ are identified with the quark mass
matrix and the chiral condensate, respectively.  Here, in the
soft-wall model, $\Sigma$ is uniquely determined for a given $M_q$ so
that Eq.~\eqref{eq:eomX0} can yield a finite and regular solution of
$X_0$.  This property is a flaw in the light-quark sector because
$M_q=0$ always leads to $\Sigma=0$ and so the spontaneous breaking of
chiral symmetry is not correctly described unless Eq.~\eqref{eq:eomX0}
is modified with higher-order potential
terms~\cite{Shock:2006gt,Gherghetta:2009ac}.  In the present work,
as we discuss later, only the heavy-quark sector is of our interest
and we need not alter Eq.~\eqref{eq:eomX0} because chiral symmetry is
largely broken in an explicit manner.
  
We must point out that the conventional soft-wall model has another
flaw in the chiral properties.  In the vicinity of the UV limit two
independent solutions of Eq.~\eqref{eq:eomX0} are definitely $z$ and
$z^3$, but if we carefully go beyond the leading order, the former
solution receives a correction by a logarithmic term as
$z\to z +(-1/2+\log z)z^3$.  This higher-order correction is small as
compared to $z$, but not small at all to another solution $z^3$.
Therefore, Eq.~\eqref{eq:eomX0} leads to a UV divergent chiral
condensate, which is an artifact of the soft-wall model.  (There is no
such logarithmic term in the hard-wall model.)  Therefore we need to
modify the model as done in
Refs.~\cite{Shock:2006gt,Gherghetta:2009ac} for example.  In the
present work we will take the following strategy.  That is, to solve
Eq.~\eqref{eq:eomX0} numerically, we will force the initial condition
by Eq.~\eqref{eq:X0-ini} and find $\Sigma$ in such a way that the
solution in the IR region contains no singularity.  This is not a
fully satisfactory resolution but is acceptable pragmatically for the
soft-wall model that is only a phenomenological model.

Let us mention on the asymptotic solutions of $X_0$ near the horizon
($z\simeq z_h$) for the finite-$T$ case.  Equation~\eqref{eq:eomX0}
can simplify by the variable change from $z$ to $t=\sqrt{3(1-z/z_h)}$,
which reduces to the Bessel equation near the horizon $t\sim0$.  It is
thus obvious that the asymptotic solutions of $X_0$ are given by the
first-kind Bessel function $J_0(t)$ which is regular and the
second-kind Bessel function $Y_0(t)$ which is divergent at $t=0$.
Since the physical solution must yield a finite action, we should pick
only $J_0(t)$ up near $z=z_h$.  To this end we need adjust an
appropriate ratio of $M_q$ and $\Sigma$ in the initial
condition~\eqref{eq:X0-ini} in the UV boundary, so that it evolves to
$J_0(t)$ near the horizon.  We will concretely carry this procedure
out in later discussions.


\section{Flavor-dependent soft-wall model}
\label{sec:flavor}

The mass spectra~\eqref{eq:Reggerho} in the soft-wall model
successfully reproduce the Regge trajectory of the light vector mesons
consisting of $u$ and $d$ quarks as seen in the previous section.  In
order to apply this model description to the heavy-quark sector, we
propose a modification on the soft-wall model in such a way that we
treat $c$ as a flavor-dependent parameter.  The following action in
our treatment is composed from two sectors; one is the light-quark
($u,d,s$) sector and the other is the heavy-quark ($c$) sector;
\begin{equation}
 S = \int d^5x \sqrt{-g}\,\tr \bigl(
  e^{-c_\rho z^2}\cL_{\text{light}} + e^{-c_{J/\psi}z^2}
  \cL_{\text{heavy}} \bigr) ,
\label{eq:action}
\end{equation}
where $\cL_{\text{heavy}}$ takes an almost identical structure with
Eq.~\eqref{eq:L} in the light-quark sector.  The only difference is
that fields in $\cL_{\text{light}}$ and in $\cL_{\text{heavy}}$ belong
to $\mathrm{U}(3)$ and $\mathrm{U}(1)$ groups, respectively.  From the
vector-meson mass formula \eqref{eq:Reggerho}, we can determine the
model parameters as
\begin{equation}
 c_\rho = 0.151\GeV^2, \qquad
 c_{J/\Psi} = 2.43\GeV^2 ,
\label{eq:crhocJPsi}
\end{equation}
to reproduce $m_\rho=0.77\GeV$ and $m_{J/\psi}=3.1\GeV$ (as we have
noted, the spectral peak is slightly different from
Eq.~\eqref{eq:Reggerho} and $c$ is shifted from the naive estimates
accordingly).  If we believe in the mass formula, the above value of
$c_{J/\psi}$ predicts the mass of the first excited state as
$4.4\GeV$, which overestimates the mass of $\psi(2S)$ that is
$3.7\GeV$.  Therefore this deviation by around $20\%$ should be taken
for a systematic error in this model~\cite{Kim:2007rt}.  This error is
of acceptable order as compared to errors associated with other
assumptions such as the large $N_c$ limit, the probe approximation,
etc.

Because of $c_\rho\ll c_{J/\psi}$, the latter term in the
action~\eqref{eq:action} is negligible to evaluate the magnitude of
$S$.  Hence, the critical temperature $T_c$ of the Hawking-Page
transition in this model is solely determined by the former term
involving $c_\rho$, that means $T_c=0.492\sqrt{c_\rho}=0.191\GeV$ is
unchanged.  In short, the bulk thermodynamics is dominated by the
former term, while the heavy-flavor sector is described by the
equation of motion deduced from the latter term the
action~\eqref{eq:action}.

Before closing this section let us comment on possible justification
of this model treatment with two scales.  One may wonder that $c$
should be common to all flavors because it is a parameter related to
the QCD string tension.  Besides, it should be more natural that
$m_{J/\psi}$ arises mostly from $M_q$ rather than $c_{J/\psi}$.  In
the soft-wall model, however, the vector-meson field has no direct
coupling with $X_0$ and so it does not depend on $M_q$.  The important
point is that $c_{J/\psi}$ as a ``renormalized'' scale can originate
from the back-reaction with the heavy charm-quark mass beyond the
probe approximation.  In fact it is pointed out in
Ref.~\cite{Shock:2006gt} that the back-reaction from the $X$ field in
the hard-wall model produces an effective soft-wall with $c$ depending
on $M_q$.  So far there is no such analysis on the back-reaction
within the framework of the soft-wall model, but it would be a
reasonable anticipation that $c$ must get larger with heavier $M_q$
once the back-reaction is taken into account.  The back-reaction
analysis in the top-down approaches~\cite{Casero:2007ae,Gursoy:2007cb}
also implies that our treatment could be pragmatically acceptable.


\section{Spectral functions}
\label{sec:spec}

In this section we proceed to actual calculation of the SPFs.  As seen
from the bulk/boundary correspondence, we can derive the SPFs in the
channel of our interest by solving the classical equation of motion in
five dimensions.  Here we take the AdS radius as $L=1$ since this
quantity disappears in the physical correlation functions.  In
addition, for convenience, we use the dimensionless energy $\omega$,
momentum $q$, and temperature $t$ in the unit of $\sqrt{c}$ and change
the variable for the fifth coordinate by $\xi=\sqrt{c}z$, so that we
can totally eliminate $c$ from the equation of motion.  Because only
$c$ is a dimensional parameter in the model, we can easily restore $c$
to discuss physical quantities.  It should be noted that spatial and
temporal components lead to distinct differential equations since
Lorentz symmetry is broken in the presence of a medium.  For the
moment we will focus on the solution of spatial fields in this
section, then in Sec.~\ref{sec:polarization} we will address a physics
insight into the dependence on the polarization direction.


\subsection{Vector Mesons}
\label{sec:vec}

Let us first consider the vector meson whose dual field is
$V_M=(A_{L,M}+A_{R,M})/2$ and then the axial-vector meson whose dual
field is $A_M=(A_{L,M}-A_{R,M})/2$.  We fix the gauge by choosing
$A_{L,z}=A_{R,z}=0$.  Besides, we impose
$\partial^\mu A_{L,\mu}=\partial^\mu A_{R,\mu}=0$ to get rid of
unphysical polarization.  The linearized equation of motion for the
spatial component $V_x$ (either $x=x_1$, $x_2$, or $x_3$) of the
vector field takes the following form,
\begin{equation}
 \p_z\Bigl[ e^{-cz^2} \sqrt{-g}\, g^{xx} g^{zz}\, (\p_z V_x) \Bigr]
  + \Bigl[ e^{-cz^2} \sqrt{-g}\, g^{xx} \p_\mu \p^\mu V_x \Bigr] = 0 .
\label{eq:eomV}
\end{equation}
Now we move to momentum space by performing the Fourier
transformation,
$V_x(x,\xi)=\int d^4x\,e^{i\sqrt{c}p\cdot x}\,\tilde{V}(p)v(\xi;p)$
and substitute the AdSBH metric~\eqref{eq:BHmetric} into
Eq.~\eqref{eq:eomV}, so that we reach,
\begin{equation}
 v'' + \biggl( \frac{3f-4}{\xi f} - 2\xi \biggr) v'
  + \biggl( \frac{\omega^2}{f^2} - \frac{q^2}{f} \biggr) v = 0 ,
\label{eq:eomv}
\end{equation}
with $p^\mu=(\omega,q^1,q^2,q^3)$ and $q^2=(q^1)^2+(q^2)^2+(q^3)^2$.
Here, as we have mentioned before, all variables are dimensionless and
the prime ($'$) stands for the derivative with respect to $\xi$.


\begin{figure}[tb]
 \includegraphics[width=0.6\textwidth]{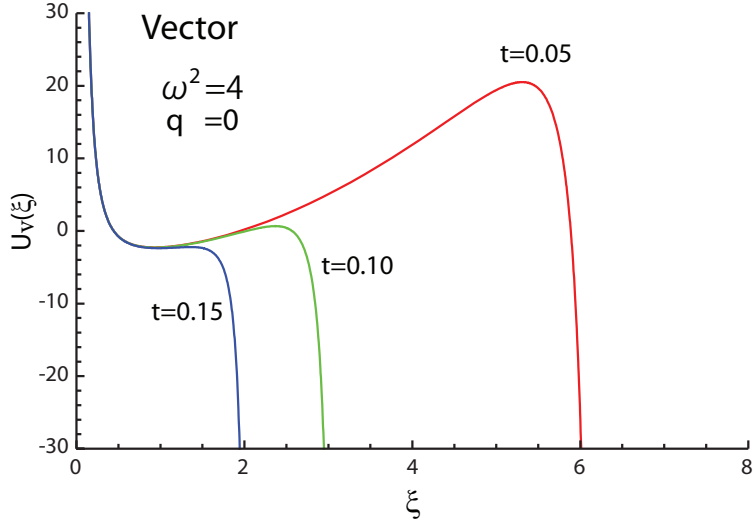}
 \caption{Potential $U_v(\xi)$ for the vector fields at dimensionless
   temperatures, $t=T/\sqrt{c}=0.05$, $0.10$, and $0.15$ at
   $\omega^2=4$ and $q^2=0$.}
 \label{fig:potenv}
\end{figure}


Before solving Eq.~\eqref{eq:eomv} it would be instructive to pursue
the analogy to the Schr\"{o}dinger equation in quantum mechanics.  The
change of the field, $u=(e^{-cz^2}\sqrt{-g}g^{xx}g^{zz})^{1/2}v$,
simplifies the equation of motion in the following form;
$u'' - U_v(\xi) u=0$ with the potential,
\begin{equation}
 U_v(\xi) = \xi^2 + \frac{3}{4\xi^2} - \frac{f'}{f}
  \biggl( 2\xi + \frac{1}{\xi} \biggr) - \frac{(f')^2}{4f^2}
  + \frac{f''}{2f} - \frac{1}{f}
  \biggl( \frac{\omega^2}{f} - q^2 \biggr) .
\label{eq:potentialv}
\end{equation}
Figure~\ref{fig:potenv} shows this potential for various dimensionless
temperatures in the unit of $\sqrt{c}$.  In the case at $T=0$ (and
thus $f=1$) the downward-convex potential, $\xi^2+3/(4\xi^2)$, yields
the discrete spectrum, $m^2=\omega^2-q^2=4n$ ($n=1,2,\dots$), only for
which the wave-function is normalizable.  We see that the higher $t$
or smaller $\xi_h=1/(\pi t)$ makes the potential less convex and
eventually it becomes monotonic at $t\simeq 0.15$.  With a monotonic
potential we cannot expect a remnant of the original spectrum any
longer.  In other words we should anticipate dissociation then.  Let
us confirm this in what follows.

At finite temperature the potential is no longer rising in the large
$z$ side and the normalizability does not quantize the spectrum.  We
can easily extract the asymptotic solutions of Eq.~\eqref{eq:eomv}
near the horizon as
\begin{equation}
 v(\xi) \;\longrightarrow\;
   c_+ \phi_+ + c_- \phi_- \qquad \text{with} \qquad
  \phi_\pm = (1-\xi/\xi_h)^{\pm i\omega\,\xi_h/4} ,
\label{eq:Asy44}
\end{equation}
in the vicinity of $\xi\to\xi_h$.  Here $\phi_+$ represents the
out-coming solution and $\phi_-$ the in-falling solution into the
black hole.  Near the origin, on the other hand, the solution has the
following asymptotic form,
\begin{equation}
 v(\xi) = A \Phi_1 + B \Phi_0 ,
\end{equation}
where $\Phi_1$ and $\Phi_0$ are two solutions of Eq.~\eqref{eq:eomV}
satisfying the following UV boundary conditions;
\begin{equation}
 \Phi_1 \;\longrightarrow\;
  -\frac{\pi}{2}\sqrt{\omega^2\!-\!q^2}\,\xi\,
  Y_1(\sqrt{\omega^2\!-\!q^2}\,\xi) ,\qquad
 \Phi_0 \;\longrightarrow\;
  \frac{2}{\sqrt{\omega^2\!-\!q^2}}\,\xi\,
  J_1(\sqrt{\omega^2\!-\!q^2}\,\xi) ,
\label{eq:UVV}
\end{equation}
around $\xi\to0$.  Here $J_1$ and $Y_1$ are the first-kind and
second-kind Bessel functions, respectively.  In the above we
normalized $\Phi_1$ and $\Phi_0$ in such a way that
$\Phi_1(\xi=\epsilon)=1$ and $\Phi_0(\xi=\epsilon)=\epsilon^2$ and
also we assumed that $\omega^2 > q^2$.  In the case that
$q^2 > \omega^2$ we should replace the above by
$\Phi_1\to \sqrt{q^2\!-\!\omega^2}\,\xi\,K_1(\sqrt{\omega^2\!-\!q^2}\,\xi)$
and
$\Phi_0\to (2/\sqrt{q^2\!-\!\omega^2})\,\xi\,I_1(\sqrt{\omega^2\!-\!q^2}\,\xi)$.
In what follows we will fix the overall normalization of $v(\xi)$ by
adopting the commonly used prescription, $A=1$, so that $B$ should be
unique once the IR boundary condition is specified.

Following the procedure elucidated in great details in
Refs.~\cite{Son:2002sd,Policastro:2002se,Teaney:2006nc} we can compute
the Green's function in Minkowskian space-time.  The IR boundary
condition must be $v(\xi\to\xi_h) = c_-\phi_-$ (i.e.\ $c_+=0$) to
acquire the retarded Green's function according to
Ref.~\cite{Son:2002sd}.  We can make $v(\xi)$ satisfy this IR boundary
condition by choosing $B$ appropriately at $\xi\simeq0$ (where $A=1$
is chosen so that $v(\xi\to\epsilon)=1$);  Then, $B$, which is now a
complex number, is uniquely fixed by the IR boundary condition;
\begin{equation}
 v(\xi) = \Phi_1(\xi) + B(\omega,q)\Phi_0(\xi)
 \;\longrightarrow\; c_- \phi_-(\xi) \qquad \text{as} \qquad
 \xi \to \xi_h .
\label{eq:vinfall}
\end{equation}
As we defined above, we can generally solve $\Phi_1$ and $\Phi_0$ from
Eq.~\eqref{eq:eomV} from the UV asymptotic forms~\eqref{eq:UVV} toward
the IR side.  If we have,
\begin{equation}
 \Phi_i(\xi) \;\longrightarrow\;
  a_i(\omega,q)\phi_+(\xi) + b_i(\omega,q)\phi_-(\xi),
\end{equation}
where $i=0,1$, then we can readily conclude $B(\omega,q)=-a_1/a_0$.


\begin{figure}[tb]
\includegraphics[width=0.6\textwidth]{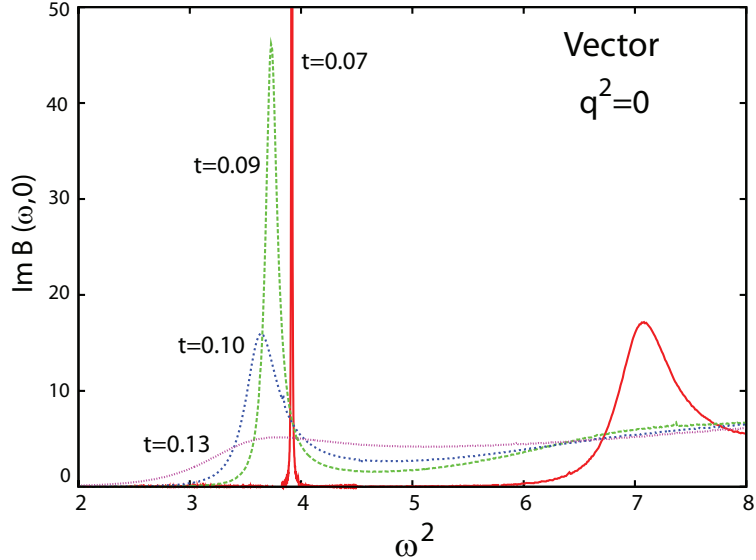}
\caption{Vector spectral functions $\mathrm{Im}\,B(\omega,0)$ for the
  temperatures, $t=0.07$, $0.09$, $0.10$, and $0.13$.  The melting
  temperature for the lowest-lying peak is about $t=0.14$.}
\label{fig:vector}
\end{figure}


Once $B$ is obtained, the bulk/boundary correspondence~\eqref{eq:GKPW}
allows us to compute the Green's function, that is given as
\begin{equation}
 D^{\text{R}}(\omega,q) = -C \lim_{\xi\to \epsilon} \biggl(
  \frac{1}{\xi} v^\ast v' \biggr)
   = -2C \biggl[  B(\omega,q) - \frac{\omega^2\!-\!q^2}{2}\,
  \ln\Bigl( \frac{e^{\gamma_E}}{2}
   \sqrt{|\omega^2\!-\!q^2|}\,\epsilon \Bigr) \biggr] ,
\label{eq:DR}
\end{equation}
where $C$ is a constant given as $C=N_c^2/(64\pi^2 L)$.  The spectral
function is, by its definition,
\begin{equation}
 \rho(\omega,q)=-\frac{1}{\pi}\mathrm{Im}D^{\text{R}}(\omega,q)
  = \frac{2C}{\pi} \mathrm{Im}B(\omega,q) .
\label{eq:spectralv}
\end{equation}
Here we note that only $B(\omega,q)$ has an imaginary part in
Eq.~\eqref{eq:DR}.

We are now ready to come to the numerical calculation.  We plot
$\mathrm{Im}B(\omega,q)$ by calculating $a_i(\omega,q)$ numerically as
a function of $\omega$ and $q$ at various temperatures and show the
SPFs at $q=0$ in Fig.~\ref{fig:vector}.  We should remark our
convention that we refer to $\mathrm{Im}B(\omega,q)$ as the SPF
neglecting an overall factor.


\subsection{Axial-vector Mesons}
\label{sec:a-vec}

Next we shall move to the SPFs in the axial-vector channel.  We can
follow exactly the same procedure as the previous one to look into the
axial-vector fields, which we denote as $A_M=(A_{L,M}-A_{R,M})/2$.
We again define the Fourier mode of the transverse component $A_x$
(where either $x=x_1$, $x_2$, or $x_3$), i.e.\ $a(\xi;p)$.  The
dimensionless equation of motion is expressed as
\begin{equation}
 a''(\xi) + \biggl( \frac{3f-4}{\xi f} - 2\xi \biggr) a'(\xi)
  + \biggl( \frac{\omega^2}{f^2} - \frac{q^2}{f} \biggr) a(\xi)
  + \frac{96\pi^2}{N_c\,\xi^2 f}\, X_0^2 \, a(\xi) = 0 .
\label{eq:eomA}
\end{equation}
We see that the above \eqref{eq:eomA} is just the same as
Eq.~\eqref{eq:eomV} for the vector fields except for the last term
involving $X_0^2$ where $X_0$ is a solution of Eq.~\eqref{eq:eomX0}.
The chiral symmetry breaking from $M_q\neq0$ and $\Sigma\neq0$ is
introduced by $X_0\sim\tfrac{1}{2}(M_q\xi+\Sigma\xi^3)$ near $\xi=0$
and is responsible for the mass splitting between the vector and
axial-vector channels.  We can also expect the last term becomes
negligible as compared to the third term for large $\omega$ or $q$, so
that the highly excited radial states exhibit degeneracy between the
vector and axial-vector mesons~\cite{Gherghetta:2009ac}, which has
been observed in the excited baryon spectrum~\cite{Cohen:2001gb}.  As
discussed in Sec.~\ref{sec:model}, the quark mass $M_q$ and the chiral
condensate $\Sigma$ are not independent in the soft-wall model and
once $M_q$ is fixed, $\Sigma$ is uniquely determined so as to yield a
regular solution of $X_0$ in the IR region under a requirement that
the UV initial condition is forced to be Eq.~\eqref{eq:X0-ini}.

Now we fix $M_q$ as the charm mass;
\begin{equation}
 M_q = M_{\text{charm}} = 1.30 \GeV ,
\end{equation} 
to derive the associated chiral condensate $\Sigma$ by the shooting
method numerically.   In our calculation we obtain
$\Sigma \simeq -(3.1\GeV)^3$, which seems overestimation but within a
reasonable range of order.  Using these $M_q$ and $\Sigma$ we can get
a regular numerical solution of the background scalar field
$X_0(\xi)$.  One noticeable fact to be mentioned is that, since $X_0$
is regular for an appropriate choice of $M_q$ and $\Sigma$ both near
the horizon $\xi\sim\xi_h$ and near the boundary $\xi=0$, the boundary
conditions for $a(\xi)$ are (almost) the same as those for $v(\xi)$ as
follows;
\begin{equation}
 a(\xi) \;\longrightarrow\; c_+ \phi_+ + c_- \phi_- ,
\label{eq:Asy-axial}
\end{equation}
near $\xi\to\xi_h$, and near the UV boundary we have
\begin{equation}
 a(\xi) = A \Phi_1' + B \Phi_0' ,
\end{equation}
where  $\Phi_1'$ and $\Phi_0'$ are two solutions of
Eq.~\eqref{eq:eomA} satisfying Eq.~\eqref{eq:UVV} with
$\omega^2-q^2$ replaced by $\omega^2-q^2+(24\pi^2/N_c)M_q^2$.


\begin{figure}[tb]
\includegraphics[width=0.6\textwidth]{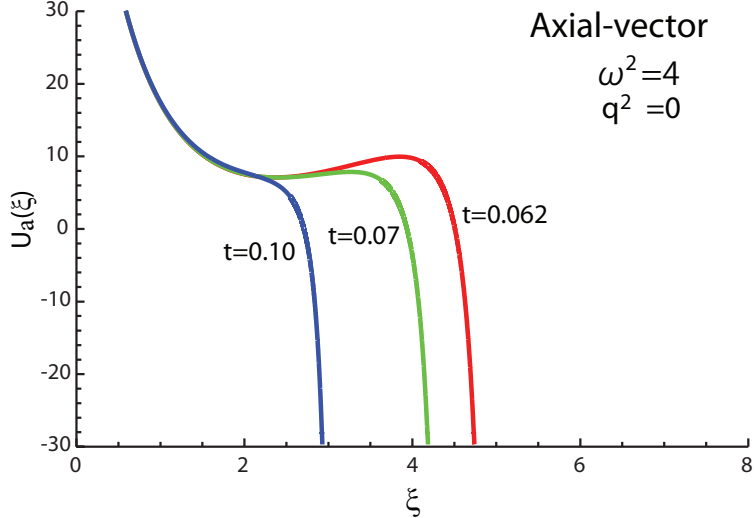}
\caption{Potential $U_a(\xi)$ for the axial-vector fields at
  dimensionless temperatures, $t=0.062$, $0.07$, and $0.10$ at
  $\omega^2=4$ and $q^2=0$.}
\label{fig:potena}
\end{figure}


Hereafter we will trace the same analysis from
Eq.~\eqref{eq:potentialv} to Eq.~\eqref{eq:spectralv} in the previous
subsection.  In the picture of the Schr\"{o}dinger equation the
corresponding potential for the axial vector case is given by
\begin{equation}
 U_a(\xi) = \xi^2 + \frac{3}{4\xi^2} - \frac{f'}{f}
  \biggl( 2\xi + \frac{1}{\xi} \biggr)
  - \frac{(f')^2}{4f^2} + \frac{f''}{2f}
  - \frac{1}{f} \biggl( \frac{\omega^2}{f} - q^2 \biggr)
  + \frac{96\pi^2}{N_c\,\xi^2f}\, X_0^2 .
\label{eq:potentiala}
\end{equation}
We show the profile of $U_a(\xi)$ in Fig.~\ref{fig:potena}.  It is
clear in view of Figs.~\ref{fig:potenv} and \ref{fig:potena} that the
axial-vector potential becomes less downward-convex earlier than the
vector case, and the shape looks monotonic already around
$t\simeq 0.10$.  Thus we can anticipate that the axial-vector spectral
peaks should melt much earlier than the vector ones.  In fact
$t\simeq 0.10$ corresponds to $T=0.10\sqrt{c}=0.16\GeV$, which is
below the deconfinement temperature $T_c=0.191\GeV$, meaning that the
axial-vector mesons should melt at the phase transition.

\begin{figure}[tb]
 \includegraphics[width=0.6\textwidth]{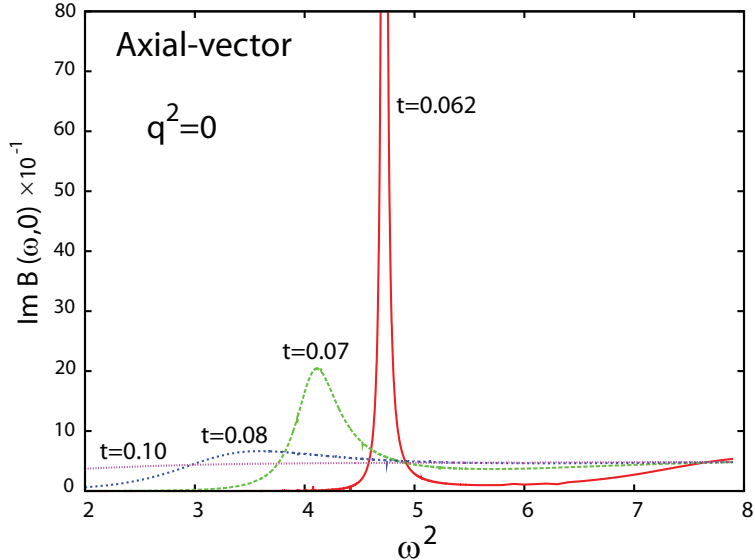}
 \caption{Axial-vector spectral functions
   $\mathrm{Im}\,B(\omega,0)\times 10^{-1}$
   for the temperatures, $t=0.062$, $0.07$, $0.08$, and $0.10$.  The
   melting temperature for the lowest-lying peak is about
   $t\simeq 0.10$.}
\label{fig:axial}
\end{figure}

Now let us derive the axial-vector SPFs.  The solution satisfying the
in-falling boundary condition determines a complex value of
$B(\omega,q)$,
\begin{equation}
 a(\xi) = \Phi_1'(\xi) + B(\omega,q) \Phi_0'(\xi) \;\longrightarrow\;
  c_- \phi_-(\xi) \qquad \text{as} \qquad \xi\to\xi_h ,
\label{eq:ainfall}
\end{equation}
Through the same procedure as elaborated in the previous subsection,
we estimate the SPFs for the axial vector mesons by evaluating
$\mathrm{Im}B(\omega,q)$ numerically.  We show our numerical results
for the axial-vector SPFs with $q=0$ in Fig.~\ref{fig:axial}.  Here we
depict $\mathrm{Im}B(\omega,q)$ divided by $10$ to make its scale
similar to Fig.~\ref{fig:vector}.  The overall factor takes a
different value depending on the vector and axial-vector channels
because of our normalization convention $A=1$.  Therefore, under the
choice of $A=1$, it is not a physically meaningful comparison to take
the absolute magnitude of spectral heights seriously.


\subsection{Discussions for Vector and Axial-vector Mesons}

Here let us make a comparison between the vector
(Fig.~\ref{fig:vector}) and axial-vector (Fig.~\ref{fig:axial})
channels.  For low temperatures, the lowest-lying peaks are located at
$\omega^2=3.92$ for the vector case and $\omega^2=4.72$ for the
axial-vector case.  If we fix $c=2.43\GeV^2$ to reproduce
$m_{J/\psi}=3.1\GeV$, then we have the mass in the axial-vector
channel as $m_{\chi_{c1}}=3.4\GeV$, that is in agreement with the
experimental value $3.51\GeV$.

It is an apparent feature seen in Figs.~\ref{fig:vector} and
\ref{fig:axial} that the spectral peaks become more collapsed and the
peak positions move smaller as $t$ increases.  We also note that the
second lowest-lying states melt far earlier than the lowest-lying
states both in the vector and axial-vector cases.  This is quite
natural because higher excited states are less stable generally.  In
terms of the potentials illustrated in Figs.~\ref{fig:potenv} and
\ref{fig:potena} a larger $\omega^2$ causes stronger absorption into
the black hole by the term, $-\omega^2/f^2$, which is negative large
near the horizon.  Furthermore, as seen from the $t=0.07$ curve in
Fig.~\ref{fig:vector}, the lowest-lying state moves smaller only
slightly, while the excited states shift more drastically.  These
qualitative properties of the SPFs are consistent with the lattice QCD
observations for the heavy quarkonia.

As seen from Figs.~\ref{fig:vector} and \ref{fig:axial}, the
dissociation takes place around $T\simeq 0.15\sqrt{c}\simeq 0.23\GeV$
for the vector lowest-lying peak and around
$T\simeq 0.10\sqrt{c}\simeq 0.16\GeV$ for the axial-vector one, where
$c=c_{J/\psi}=2.43\GeV^2$ as discussed before.  The deconfinement
transition occurs at $T_c=0.492\sqrt{c}=0.19\GeV$ as mentioned in
Sec.~\ref{sec:flavor}.  Thus, in our soft-wall QCD model, the vector
charmonium $J/\psi$ melts above the critical temperature;
$T\simeq 1.2T_c$, while the dissociation temperature of the
axial-vector charmonium $\chi_{c1}$ is much lower; $T\simeq 0.8T_c$,
which indicates that $\chi_{c1}$ does not survive above $T_c$ and
melts suddenly at the deconfinement transition.
It is obvious in our argument that the chiral symmetry breaking
induced by $X_0$ causes this difference between the vector and
axial-vector SPFs.


\subsection{Scalar and Pseudo-scalar Mesons}

Here we go on to the SPF for the scalar and pseudo-scalar fields,
whose lowest-lying peak can be identified as $\chi_{c0}$ and
$\eta_{c0}$.  As discussed in Sec.~\ref{sec:model}, we can introduce
the dual fields of the scalar and pseudo-scalar mesons by decomposing
the bi-fundamental scalar field as $X=e^{2i\Pi(x,z)}[X_0(z)+S(x,z)]$,
where $X_0$ is the background part, $S$ is the scalar field and $\Pi$
the pseudo-scalar field.  We will denote the Fourier modes of $S$ and
$\Pi$ as $s$ and $\pi$, respectively, in what follows below.

Before addressing the SPFs, we need to consider the holographic
renormalization and counter
terms~\cite{de Haro:2000xn,Skenderis:2002wp,Karch:2005ms} to give
regular results near the boundary as well as physically meaningful SPFs
in the scalar channel.  The action with respect to the scalar and
pseudo-scalar fields in the quadratic order of $S^2$ and $\Pi^2$ is
given by
\begin{equation}
 S = \int_{z=0} d^4x \, \frac{e^{-cz^2}}{z^3} \bigl(
  - S'S - 4X_0^2 \Pi'\Pi \bigr) + S_{\text{eom}}
\end{equation}
after the integration by parts, in which the functional derivative of
$S_{\text{eom}}$ leads to the equations of motion.  The first term is
UV divergent at $z\to0$ and requires the renormalization counter term
that is constructed in such a way that the covariance holds;
\begin{equation}
 S_{\text{ren}} = \int_{z=0} d^4x e^{-cz^2}\sqrt{-\gamma}\, X^{2} ,
\label{eq:counter}
\end{equation}
where $\gamma$ is the determinant of the induced metric defined as
$\gamma_{\mu\nu}=\diag(-fz^{-2},z^{-2},z^{-2},z^{-2})$ and thus
$\sqrt{-\gamma}\sim z^{-4}$ near $z=0$.  The renormalized action is
defined as
$S+S_{\text{ren}}$~\cite{de Haro:2000xn,Skenderis:2002wp,Karch:2005ms}.
Once we comply with this renormalized procedure, we can follow the
same procedure as in the previous case for the vector and axial-vector
mesons.  In this way we find the dimensionless equation of motion for
the scalar and pseudo-scalar fields;
\begin{align}
 & s''(\xi) + \biggl( \frac{f-4}{\xi f} - 2\xi \biggr) s'(\xi) +
  \biggl( \frac{\omega^2}{f^2} - \frac{q^2}{f} +
  \frac{3}{\xi^2 f} \biggr) s(\xi) = 0 ,
\label{eq:eoms} \\
 & \pi''(\xi) + \biggl( \frac{f-4}{\xi f} - 2\xi
  + \frac{2X_0'}{X_0} \biggl) \pi'(\xi) +
  \biggl( \frac{\omega^2}{f^2} - \frac{q^2}{f} \biggr) \pi(\xi) = 0 .
\label{eq:eompi}
\end{align}
Here the dependence on the quark mass and the chiral condensate is
introduced into the pseudo-scalar solution through the second
(first-derivative) term in Eq.~\eqref{eq:eompi}, while the scalar
equation of motion does not have such a term.  This difference should
be attributed to distinct mass spectra and dissociation temperatures
between the scalar and pseudo-scalar mesons like the vector and
axial-vector cases.


\begin{figure}[tb]
 \includegraphics[width=0.6\textwidth]{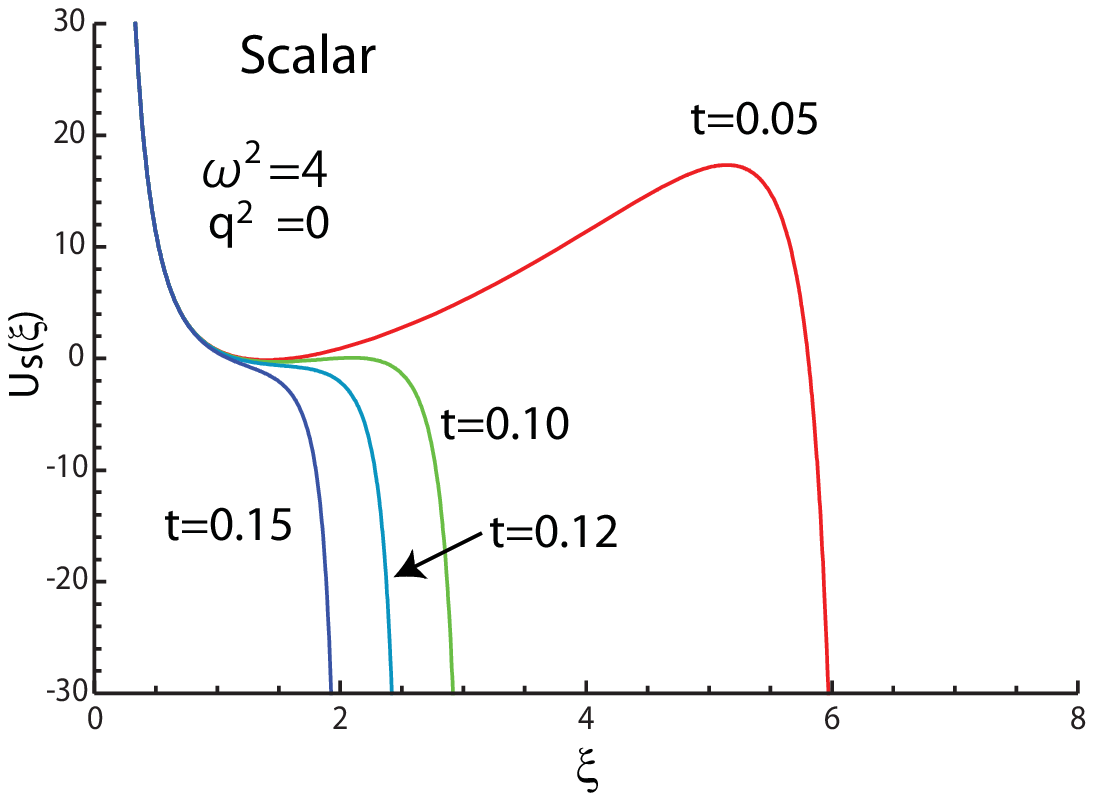}
 \caption{Potential $U_s(\xi)$ for the scalar fields at dimensionless
   temperatures, $t=0.05$, $0.10$, $0.12$, and $0.15$ at $\omega^2=4$
   and $q^2=0$.}
\label{fig:potens}
\vspace{3mm}

 \includegraphics[width=0.6\textwidth]{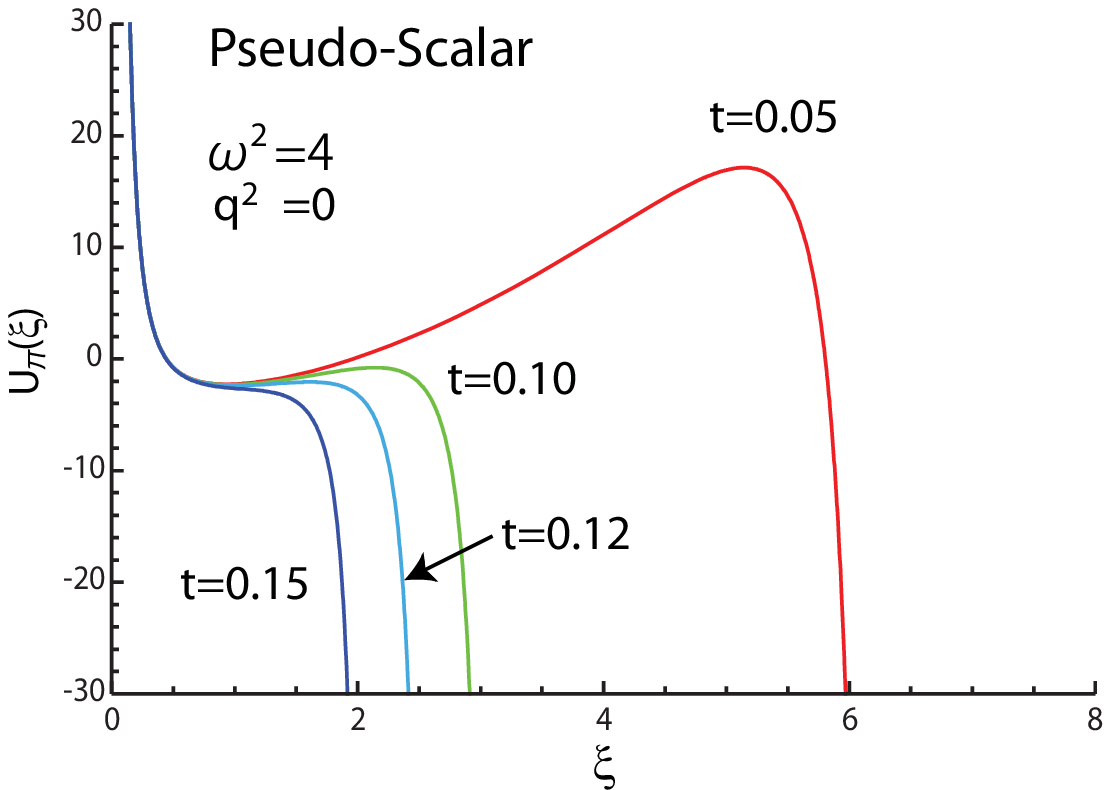}
 \caption{Potential $U_\pi(\xi)$ for the pseudo-scalar fields at
   dimensionless temperatures, $t=0.05$, $0.10$, and $0.15$ at
   $\omega^2=4$ and $q^2=0$.}
\label{fig:potenp}
\end{figure}


Here again, we shall trace the same procedures as those from
Eq.~\eqref{eq:potentialv} to Eq.~\eqref{eq:spectralv}.  The potentials
for the scalar and pseudo-scalar fields in the picture of the
Schr\"{o}dinger equation are read from the equations of motion as
\begin{align}
 U_s(\xi) &= \xi^2 + \frac{15}{4\xi^2}
  - \frac{f'}{2f} \biggl( 2\xi + \frac{3}{\xi} \biggr)
  - \frac{(f')^2}{4f^2} + \frac{f''}{2f}
  - \biggl( \frac{\omega^2}{f^2} - \frac{k^2}{f} \biggr) ,\\
 U_\pi(\xi) &= \xi^2 + \frac{15}{4\xi^2}
  - \frac{f'}{2f} \biggl( 2\xi + \frac3\xi \biggr)
  - \frac{(f')^2}{4f^2} + \frac{f''}{2f}
  - \biggl( \frac{\omega^2}{f^2} - \frac{k^2}{f} \biggr)
  - \frac{3}{\xi^2f} .
\label{eq:pipoten}
\end{align}
Remarkably, the potential for the pseudo-scalar field is independent
of the background solution $X_0$ since all the terms depending on
$X_0$ are put together into a form of the equation of
motion~\eqref{eq:eomX0}.  The difference between the scalar and
pseudo-scalar fields is only the last term in Eq.~\eqref{eq:pipoten}.
We depict these potentials in Figs.~\ref{fig:potens} and
\ref{fig:potenp}.  The results look very similar and turn monotonic
around $t\simeq 0.12$.

Then we find that the behavior of the solutions in the near-horizon
region is again given by $\phi_\pm$.  In the opposite side of the UV
limit we have two solutions for the scalar and pseudo-scalar fields.  
That is,
\begin{equation}
  s(\xi) = A_s\Phi_1'' + B_s \Phi_0'' ,\qquad
 \pi(\xi) = A_\pi \Phi_1 + B_\pi \Phi_0 ,
\end{equation}
where $\Phi_0$ and $\Phi_1$ are defined in Eq.~\eqref{eq:UVV} and
$\Phi_0''$ and $\Phi_1''$ are the solutions of the equation of
motion~\eqref{eq:eoms} with the boundary conditions;
$\Phi_0''(\xi=\epsilon)=\epsilon$ and
$\Phi_1''(\xi=\epsilon)=\epsilon^3$.

Here let us note that, strictly speaking, the scalar field
corresponding to the scalar source at the boundary is $s(\xi)/\xi$,
and thus the boundary solutions behave asymptotically as $1$ and
$\xi^2$ like the other channels.  Therefore the SPFs are characterized
by the imaginary part of the complex coefficients $B_s$ and $B_\pi$.
To calculate the retarded Green's function we fix $B_s$ and $B_\pi$
requiring the in-falling boundary condition near the horizon.


\begin{figure}[tb]
\includegraphics[width=0.6\textwidth]{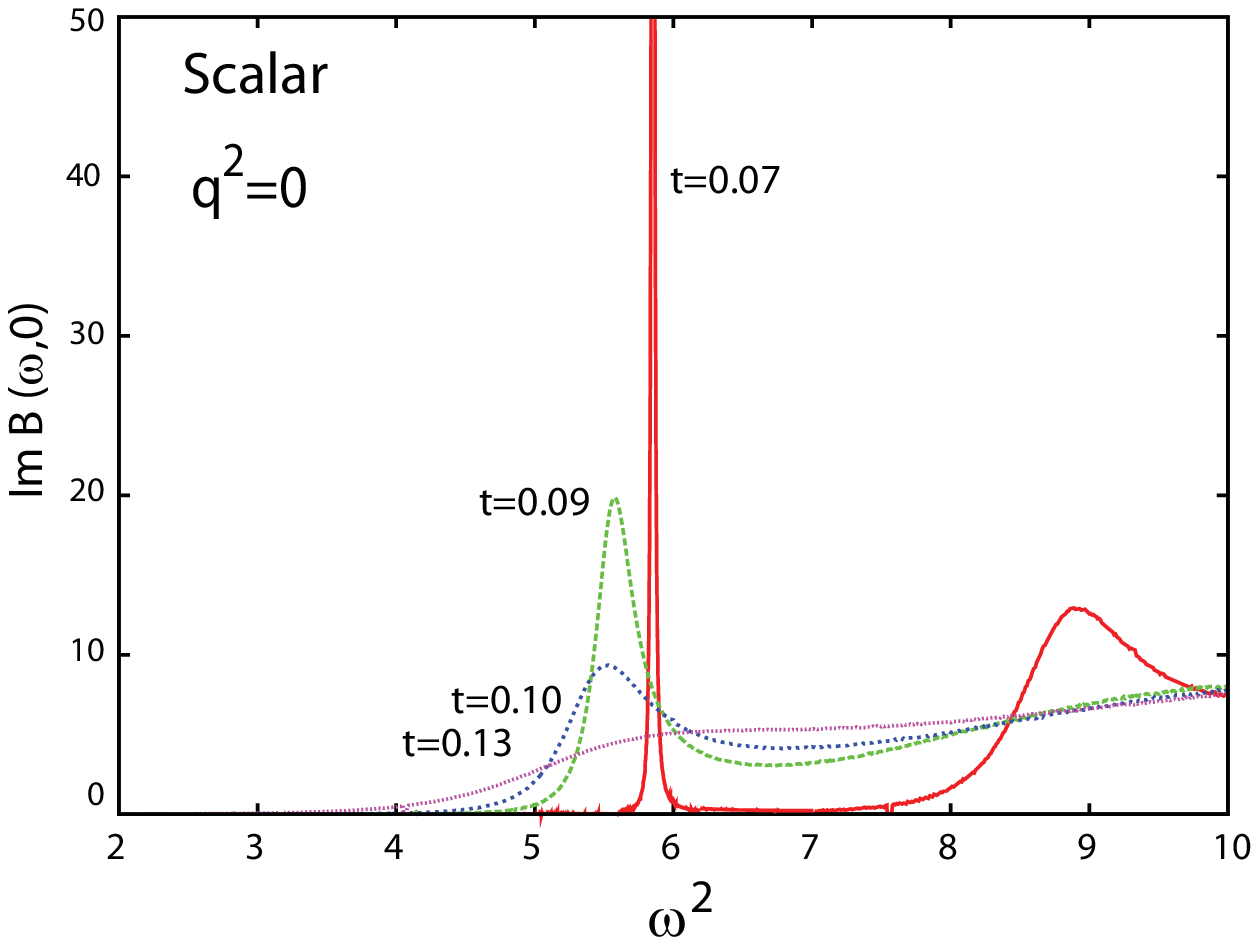}
\caption{Scalar spectral functions $\mathrm{Im}\,B(\omega,0)$ for
  the temperatures, $t=0.07$, $0.09$, $0.10$, and $0.13$.}
\label{fig:scalar}
\vspace{3mm}

\includegraphics[width=0.6\textwidth]{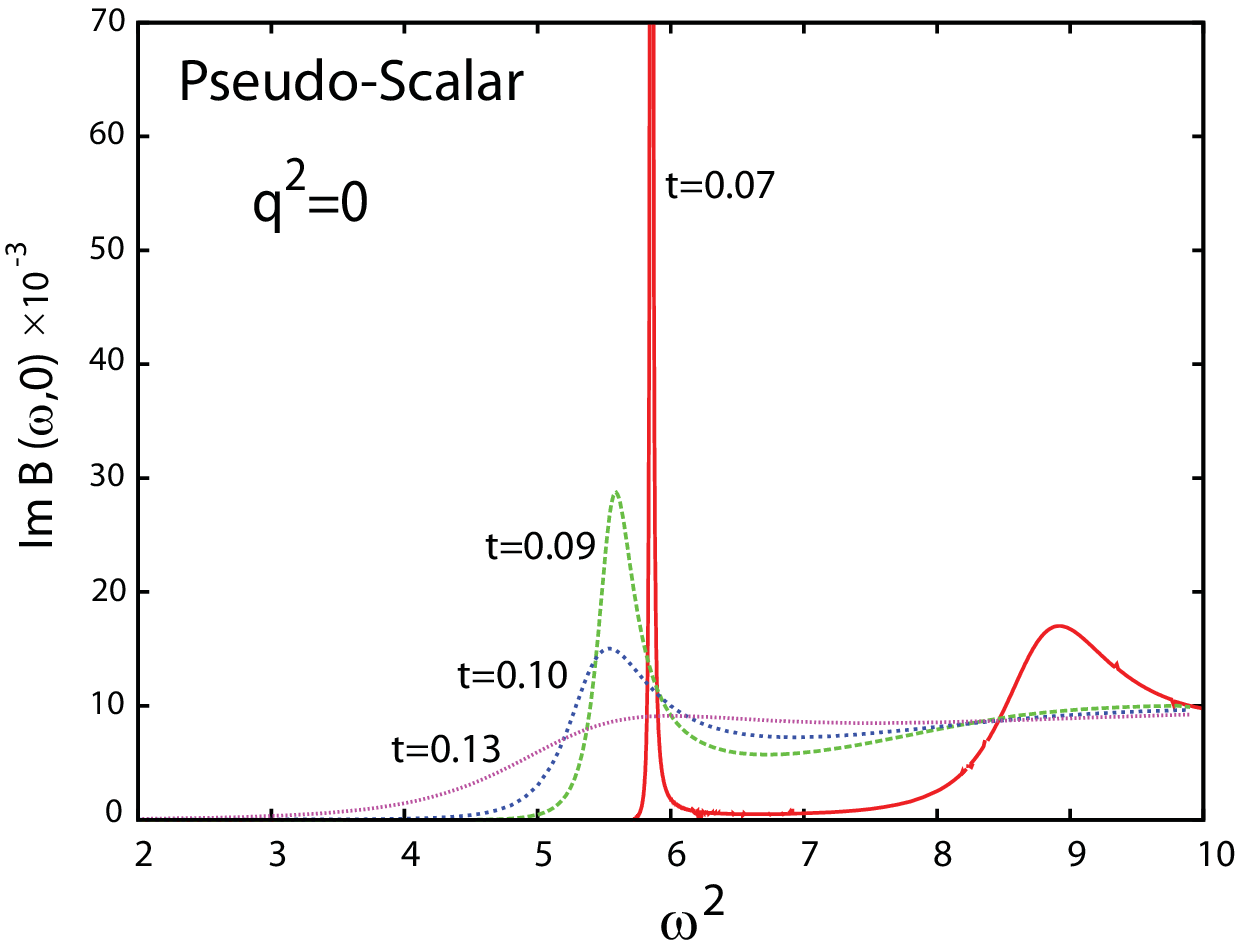}
\caption{Pseudo-scalar spectral functions
  $\mathrm{Im}\,B(\omega,0)\times10^{-3}$
  for the temperatures, $t=0.07$, $0.09$, $0.10$, and $0.13$.}
\label{fig:pseudo}
\end{figure}


We numerically calculate $\mathrm{Im}B_s$ and $\mathrm{Im}B_\pi$ and
make plots as a function of $\omega$ at $q=0$ in
Figs.~\ref{fig:scalar} and \ref{fig:pseudo}.  Here we present
$\mathrm{Im}B_\pi$ divided by $10^3$ since the normalization $A=1$
gives an irrelevant overall factor again.  The lowest-lying peak at
small temperature for the scalar and pseudo-scalar channels is located
at $\omega^2=5.85$.  The model outputs for the meson masses are thus
$m_{\chi_{c0}}\approx m_{\eta_c}=3.8\GeV$, which is not good as
compared with $m_{\eta_c}=3.0\GeV$ and $m_{\chi_{c0}}=3.4\GeV$.  The
scalar spectral peaks are located around $\omega^2=(4n+6)c$ with
$n=0,1,2,\dots$ for low temperatures, as is consistent with the
results in Ref.~\cite{Colangelo:2009ra}.  Here we see that the
pseudo-scalar spectral peaks are found at nearly the same positions as
the scalar mesons.

The lowest-lying peaks are gradually collapsed and moving smaller as
the temperature increases, while the excited peaks dissociate much
earlier and shift more drastically.  These spectral patterns are
qualitatively similar to the vector and axial-vector cases.  We can
observe that the lowest-lying spectral peaks melt out around
$t\simeq 0.13$, i.e.\ $T=0.13\sqrt{c}=0.20\GeV$ for the scalar and
pseudo-scalar channels both, which is slightly above the deconfinement
temperature; $T\simeq 1.05T_c$.  If we take a closer look at the
respective SPFs, we notice that the scalar meson melts only slightly
earlier than the pseudo-scalar meson.  The difference is, however,
hardly perceivable and we can say that the scalar and pseudo-scalar
channels are degenerate regardless of the chiral symmetry breaking.

\section{More Discussions on the Vector Mesons}

We have seen that only the vector meson, i.e.\ $J/\psi$, survives
above $T_c$ (up to $T\simeq 1.2T_c$ in our model), the axial-vector
$\chi_{c1}$ suddenly disappears at $T=T_c$, and the scalar $\chi_{c0}$
and pseudo-scalar $\eta_c$ immediately melt around $T\simeq 1.05T_c$.
Therefore, it should be worth while taking a more serious look at the
vector SPFs only.  In this section we analyze the vector SPFs by
deducing the relation between the mass shift $\Delta m$ and the width
broadening $\Gamma$ with changing $t$.  We also discuss the evolution of
the SPFs at finite momentum $q$.  Finally we briefly mention on the
dependence on the polarization direction.


\subsection{Mass Shift and Width Broadening}


\begin{figure}
\includegraphics[width=0.7\textwidth]{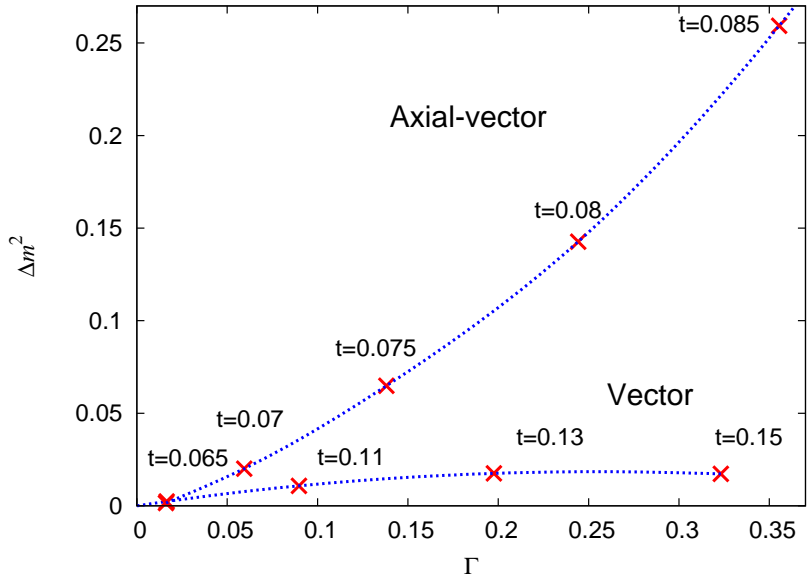}
 \caption{Mass shift squared as a function of the width with changing
  temperatures.  The dashed curve smoothly connects the calculated
  points at various temperatures for the vector and axial-vector
  channels.}
\label{fig:mass}
\end{figure}


According to our previous work~\cite{Fujita:2009wc}, a functional
ansatz, $a\,\omega^b/[(\omega-\omega_0)^2+\Gamma^2]$ can fit the SPFs
pretty well.  In this way we can numerically read the peak position
$\omega_0(t)$ (leading to the mass shift defined by
$\Delta m(t)=\omega_0(0)-\omega_0(t)$) and the width $\Gamma(t)$
determined as a function of $t$.

In Fig.~\ref{fig:mass} we plot the mass shift squared $(\Delta m)^2$
and the width $\Gamma$ associated with the lowest-lying peak in the
vector and axial-vector SPFs.  It is an intriguing finding from
Fig.~\ref{fig:mass} that, even though the SPFs shown in
Figs.~\ref{fig:vector} and \ref{fig:axial} look similar at a glance,
the qualitative behavior of the mass shift is completely different.
In the case of the vector meson $\Delta m^2$ is saturated as $t$
increases, while $\Gamma$ continues growing.  The relation between
$\Delta m$ and $\Gamma$ has been investigated in the QCD sum
rule~\cite{Morita:2007pt}, which is seemingly inconsistent with
Fig.~\ref{fig:mass} but a careful consideration clarifies
consistency~\cite{Morita}.  In the previous work in
Ref.~\cite{Fujita:2009wc} we proposed a definition for the
dissociation temperature by means of the saturating behavior of
$\Delta m^2$ around $t=0.14$.  This working definition works for the
vector meson, while the axial-vector peak keeps becoming lighter
(i.e.\ larger $\Delta m^2$) and broader (i.e.\ larger $\Gamma$) and
thus there is no saturation observed.  It is an interesting question
whether our prediction about the relation between $\Delta m^2$ and
$\Gamma$ in the axial-vector channel can be confirmed or not in other
models such as the QCD sum rule.


\subsection{Finite Momentum}

In this subsection we briefly discuss the momentum dependence of the
SPFs.  There are several lattice QCD results on the $J/\psi$ SPFs at
finite momentum~\cite{Datta:2004js,Aarts:2006cq}.  Although it is not
clear whether the lattice simulation achieves accuracy enough to be
reliable, the general tendency is that the spectral peaks are
attenuated as $q$ get larger.

\begin{figure}
\includegraphics[width=0.7\textwidth]{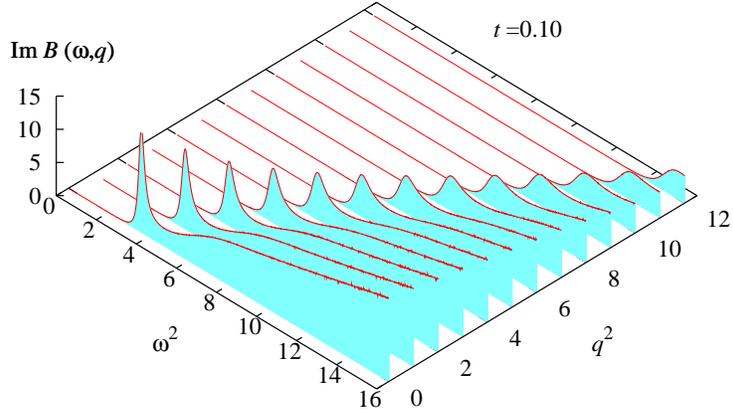}
\caption{Spectral functions $\mathrm{Im}B(\omega,q)$ as a function of
  $\omega$ and $q$ for a fixed temperature $t=0.10$, where $\omega$,
  $q$, and $t$ are all dimensionless in the unit of $\sqrt{c}$.}
\label{fig:spect010}
\end{figure}

Here we present the results only for the vector channel because only
$J/\psi$ survives above $T_c$ in the soft-wall model, which is our
finding in this paper.  We plot the numerical results in
Fig.~\ref{fig:spect010} for $q^2$ ranging from $0$ to $12$ with
$t=0.10$ fixed.  We choose this temperature to make it easier to grasp
the qualitative feature of the fairly prominent peaks in the SPFs,
though we know that $t=0.10$ is below $T_c$.  The conclusion is, of
course, unaltered even if we carry the analysis out on the case at
$T>T_c$ as long as the peak remains.

It is apparent that the spectral peak is gradually collapsed as $q$
increases.  This result is quite non-trivial and peculiar to the
non-perturbative regime since in the perturbative evaluation a larger
$q$ makes the spectral peak less sensitive to the medium
effect~\cite{Hidaka:2002xv}.  It has been studied in
Refs.~\cite{Liu:2006nn} that, in the strongly-correlated $N=4$ Super
Yang-Mills theory, $J/\psi$ melts at high $q$, or in a frame where
$J/\psi$ is at rest, it melts under the hot wind of QGP matter.  This
conclusion has been confirmed in the top-down holographic QCD model
later~\cite{Faulkner:2008qk}.  The discussions in
Refs.~\cite{Liu:2006nn,Faulkner:2008qk} did not originate from the
shape of the SPFs, however.  Our present results add another
confirmation of the hot screening scenario, and maybe the first
evidence directly inferred from the shape of the SPFs.


\subsection{Polarization Dependence}
\label{sec:polarization}

We finally discuss the vector solution of the equation of
motion~\eqref{eq:eomv}.  We can easily see that the equation of motion
takes slightly different forms depending on whether the polarization
is $V_x$ or $V_0$.  One may think that this difference should be
interpreted as distinct behavior of the transverse and longitudinal
modes in a medium.  In fact, usually, if we see some vector-like
collective excitations with finite momentum $q$ that is directed to
$q_3$ for the moment, a linear combination of 0th and 3rd components
describes the longitudinal mode, which becomes distinguishable from
the transverse modes.

In this sense, it is surprising that Eq.~\eqref{eq:eomv} takes
completely the same form regardless of the choice of $x=x_1$, $x_2$,
or $x_3$, that is, Eq.~\eqref{eq:eomv} is insensitive to whether
$V_\mu$ is parallel or perpendicular to $q$.  This is a very strong
statement.  Usually, the transverse and longitudinal polarizations
become degenerated only when $q=0$ because the rotational symmetry is
restored then~\cite{Kovtun:2005ev}.  (See also Ref.~\cite{Kim:2006ut}
for the hydrodynamic limit of the longitudinal mode).

We point out that the above-mentioned statement can translate into the
interpretation that there is no jet in the strong coupling
regime~\cite{Hatta:2008tx}.  In our case the equation of motion and
thus all the physical results are given as a function of $q^2$ only
and $V_x$'s are completely equivalent for $x=x_1$, $x_2$, and $x_3$.
This means that the rotational symmetry of the system is not broken at
all even though a finite momentum $q$ is inserted.  One can
intuitively understand this as quick realization of the equipartition
of inserted momentum over the system.  Such a picture is very similar
to the finding of Ref.~\cite{Hatta:2008tx} that no jet can exist in a
strong-coupling $\mathcal{N}=4$ supersymmetric Yang-Mills medium.
Actually, if we assume the vector dominance, we can apply our results
of the vector SPFs directly for the problem of the dilepton
production, which may be an interesting direction regarding the future
extension of our work.


\section{Summary}
\label{sec:summary}

In this paper we derived the SPFs of meson states in the vector,
axial-vector, scalar, and pseudo-scalar channels at finite temperature
using the soft-wall AdS/QCD model.  We pointed out that the SPFs in
these channels have several qualitative features as follows:
\vspace{1mm}

(i) Only $J/\psi$ survives above the deconfinement transition up to
$T\simeq 1.2T_c$ and $\chi_{c1}$ completely melts at the transition.
The scalar $\chi_{c0}$ and pseudo-scalar $\eta_c$ are almost
degenerate in our model and melt soon above $T_c$.
\vspace{1mm}

(ii) The relation between the mass shift squared $\Delta m^2$ and the
width $\Gamma$ is characteristic to $J/\psi$ and $\chi_{c1}$.  In the
vector channel $\Delta m^2$ looks almost linearly proportional to
$\Gamma$ at small temperatures until it is saturated at the
dissociation.  In the axial-vector channel, in contrast, both
$\Delta m^2$ and $\Gamma$ keep growing up with increasing
temperature.
\vspace{1mm}

(iii) The spectral peaks diminish at finite momentum, as is consistent
with the scenario of the $J/\psi$ suppression under a hot wind of QGP
matter.
\vspace{1mm}

(iv) All the results on the vector and axial-vector meson properties
respect the rotational symmetry regardless of the presence of the
momentum insertion.  This should be interpreted as the equipartition
of the momentum in a medium in the strong-coupling regime.
\vspace{1mm}

For more realistic studies to investigate the non-perturbative aspect
of QCD, we need to construct a better model than the soft-wall QCD
model that we adopted in this work.  In the process of concrete
computations, in fact, we realized that the conventional soft-wall
model does not satisfy the requirement that the bi-fundamental scalar
field should be $X_0\sim M_q z + \Sigma z^3$ near the UV boundary
($z\sim 0$) but leads to a logarithmic correction $z^3\log(z)$.  The
presence of $z^3\log(z)$ in the solutions of the equation of motion
brings huge uncertainty in evaluating the chiral condensate $\Sigma$
numerically for a given quark mass $M_q$.

In addition to this problem of the asymptotic solution, there is
another problem, that is, the conventional soft-wall model cannot
describe the chiral phase transition.  In reality what should be
expected is that chiral symmetry is restored at the deconfinement
transition simultaneously and then the vector and axial-vector
channels become identical.  In our case the quark mass is
significantly heavy and breaks chiral symmetry badly, and thus we can
consider that the lack of chiral restoration does not affect our
results.  Nevertheless, it is not clear a priori if not only $M_q$ but
also $\Sigma$ have a substantial effect on the discrepancy between the
vector and axial-vector mesons.  To circumvent all these problems we
will be able to use the modified soft-wall
model~\cite{Shock:2006gt,Gherghetta:2009ac} or the top-down approaches
such as the D3/D7 and Sakai-Sugimoto models~\cite{Sakai:2004cn}.

There are many directions in which the present work can be extended in
the future.  One example is the application to the dilepton production
problem for which the vector SPF is the essential ingredient.  We
could maybe use more realistic holographic models mimicking the QCD
equation of state~\cite{Gubser:2008ny}.  It is also an interesting
generalization to introduce not only the temperature effect but also
the baryon density or the baryon chemical potential.  Then, the
Chern-Simons coupling mixes the vector and axial-vector
mesons~\cite{Domokos:2007kt}, which leads to an additional spectral
broadening~\cite{Harada:2009cn}.


\begin{acknowledgments}
We are grateful to Yoshimasa Hidaka, Youngman Kim, and
Misha Stephanov for discussions.  M.~F.\ thanks Hiroyuki Hata for
useful advice.  T.~M.\ and M.~M.\ thank Noriaki Ogawa for technical
supports.  K.~F.\ is supported by Japanese MEXT grant No.\ 20740134
and also supported in part by Yukawa International Program for Quark
Hadron Sciences.
T.~K.~(No.\ 21-951), T.~M.~(No.\ 21-1226) and M.~M.~(No.\ 21-173) are supported by Grand-in-Aid for the Japan Society for Promotion of Science (JSPS) Research Fellows.  
\end{acknowledgments}


\end{document}